\documentclass[aps,prb,twocolumn,superscriptaddress,showpacs,floatfix]{revtex4}
\usepackage{graphicx}

\begin{document}

\title{Non-equilibrium spin polarization effect in spin-orbit coupling
system and contacting leads}
\author{Yongjin Jiang}
\email{jyj@zjnu.cn}
\affiliation{Physics department of ZheJiang Normal University, Jinhua, Zhejiang Province,
321004,P.R.China}

\begin{abstract}

We study theoretically the current-induced spin polarization
effect in a two-terminal mesoscopic structure which is composed of
a semiconductor two-dimensional electron gas (2DEG) bar with
Rashba spin-orbit (SO) interaction and two attached ideal leads.
The nonequilibrium spin density is calculated by solving the
scattering wave functions explicitly within the ballistic
transport regime. We found that for a Rashba SO system the
electrical current can induce spin polarization in the SO system
as well as in the ideal leads. The induced polarization in the
2DEG shows some qualitative features of the intrinsic spin Hall
effect. On the other hand, the nonequilibrium spin density in the
ideal leads, after being averaged in the transversal direction, is
independent of the distance measured from the lead/SO system
interface, except in the vicinity of the interface. Such a lead
polarization effect can even be enhanced by the presence of weak
impurity scattering in the SO system and may be detectable in real
experiments.

\end{abstract}

\maketitle


\section{Introduction}

Recently there has been hot research interest in spin-polarized transport
phenomena in semiconductor systems with spin-orbit ( SO ) coupling because
of their potential applications in the field of semiconductor spintronics%
\cite{Wolf2001,semibook2002,Dassarma2004}. One of the principal
challenges in semiconductor spintronics is the efficient injection
and effective manipulation of non-equilibrium spin densities
and/or spin currents in nonmagnetic semiconductors by electric
means, and within this context, the recently predicted phenomenon
of \textit{intrinsic spin Hall effect} would be much attractive.
This phenomenon was conceived to survive in some semiconductor
systems with intrinsic spin-orbit interactions and it consists of
a spin current contribution in the direction perpendicular to a
driving charge current circulating through a sample. This
phenomenon was firstly proposed to survive in \textit{p}-doped
bulk semiconductors with spin-orbit split band structures by
Murakami et al.\cite{MurakamiScience2003} and in two-dimensional
electronic systems with Rashba spin-orbit coupling by Sinova et
al..\cite{SinovaPRL2004} In the last two years lots of theoretical
work have been devoted to the study of this phenomenon and
intensive debates
arose on some fundamental issues concerning this effect. \cite%
{JPHu2003,RashbaPRB2003,SQShen2004,InouePRB2004,MishchenkoPRL2004,RashbaPRB2004,LShengPRL2005,MurakamiPRB2004,Bernevig2004,Nikolic2004,Nikolic2005,yqli,PZhang2005,BernevigB2004,DNSheng2005}
Now it was well established that this effect should be robust against
impurity scattering in a mesoscopic semiconductor system with intrinsic
spin-orbit coupling, but it cannot survive in a diffusive two-dimensional
electron gas with k-linear intrinsic spin-orbit coupling due to vertex
correction from impurity scattering.\cite%
{InouePRB2004,MishchenkoPRL2004,RashbaPRB2004,DNSheng2005}. It also should
be noted that, prior to the prediction of this effect, a similar phenomenon
called \textit{extrinsic spin Hall effect} had also been predicted by
Dyakanov-Perel\cite{MD} and Hirsch\cite{Hirsch99,Zhang00}. Unlike the
intrinsic spin Hall effect, the extrinsic spin Hall effect arises from
spin-orbit dependent impurity scattering and hence it is a much weak effect
compared with the intrinsic spin Hall effect, and especially, it will
disappear completely in the absence of spin-orbit dependent impurity
scattering.

A physical quantity which is intuitively a direct consequence of a spin Hall
effect in a semiconductor strip is the nonequilibrium spin accumulation near
the transversal boundaries of the strip when a longitudinal charge current
circulates through it. The recent experimental observations in \textit{n}%
-doped bulk GaAs\cite{KatoScience2004} and\emph{\ }in \textit{p}-doped
two-dimensional GaAs\cite{WunderlichPRL2005} have revealed the possibility
of generating non-equilibrium edge spin accumulation in a semiconductor
strip by an extrinsic\cite{KatoScience2004} or intrinsic\cite%
{WunderlichPRL2005} spin Hall effect, though from the theoretical points of
view some significant controversies may still exist for the detailed
theoretical interpretations of some experimental data.\cite%
{BernevigB2004,Nikolic2004,Usaj2005,Zyang2005,jianghu1}. It also
should be noticed that, in addition to the spin Hall effect, there
may be some other physical reasons that may lead to the generation
of electric-current-induced spin polarization in a spin-orbit
coupled system. For example, the experimental results reported in
Ref\cite{Silov} had demonstrated another kind of
electric-current-induced spin polarization effect in a spin-orbit
coupled
system, which can be interpreted as the inverse of the photogalvanic effect%
\cite{Silov}. In the present paper we are interested in the question that,
besides the transverse boundary spin accumulations, what kinds of other
physical consequences can be predicted for a semiconductor strip with
intrinsic SO coupling. To be specific, we will consider a strip of a
mesoscopic two-dimensional electron gas with Rashba spin-orbit coupling
connected to two ideal leads. For such a system, we show that when a
longitudinal charge current circulates through the strip, besides the
transverse boundary spin accumulations in the strip, non-equilibrium spin
polarizations can also be induced in the leads, and the non-equilibrium spin
polarizations in the leads will also depend sensitively on the spin-orbit
coupling parameters in the strip. Furthermore we will show that in the
absence of spin-orbit coupling in the leads, the non-equilibrium spin
polarizations in the leads will be independent of the distance away from the
border between the lead and the strip, except in the vicinity of the border
where local states play an important role. The signs and the amplitudes of
the non-equilibrium spin polarization in the leads is adjustable by
switching the charge current direction and by tuning the Rashba SO coupling
strength in the strip. Furthermore, in the presence of weak disorder inside
the SO strip, the spin polarization effect in the leads can be even \emph{%
enhanced} while spin polarization in the strip is decreased, so such an
effect might be more easily detectable than the spin polarization inside the
strip. The non-equilibrium lead polarization effect provides a new
possibility of all-electric control on spin degree of freedom in
semiconductor and thus have practical application potential.

The paper will be organized as following: In section II the model
Hamiltonian and the theoretical formalism used in the calculations will be
introduced. In section III some symmetry properties and numerical results
about the non-equilibrium spin polarizations both in the strip and in the
leads will be presented and discussed. At last, we summarize main points of
the paper in section IV.

\begin{figure}[tbh]
\includegraphics{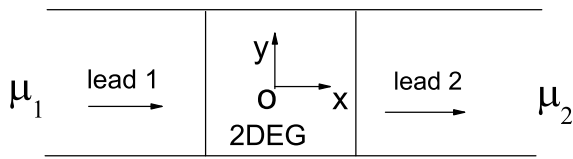}
\caption{schematic geometry of multi-terminal setup.}
\label{fig:geometry}
\end{figure}

\section{Method of solving scattering wave function and multi-terminal
scattering matrix}

The system that we will consider in the present paper is depicted in Fig.~%
\ref{fig:geometry}, which consists of a strip of a mesoscopic Rashba
two-dimensional electron gas connected to two half-infinite ideal leads.
Each lead is connected to an electron reservoir at infinity which has a
fixed chemical potential. In the tight binding representation, the
Hamiltonian for the total system reads:

\begin{eqnarray}
\lefteqn{H=-t\sum_{p}\sum_{<i,j>\sigma}(C_{p_{i}\sigma}^{\dag}C_{p_{j}%
\sigma}+h.c.)+\sum_{R_{i},\sigma}w_{R_{i}}C_{R_{i}\sigma}^{\dag}C_{R_{i}%
\sigma}}  \nonumber \\
&&-t\sum_{<R_{i},R_{j}>\sigma}(C_{R_{i}\sigma}^{\dag}C_{R_{j}\sigma}+h.c.)
\nonumber \\
&&-t_{R}\sum_{R_{i}}[i(\Psi_{R_{i}}^{\dag}\sigma^{x}\Psi_{R_{i}+y}-%
\Psi_{R_{i}}^{\dag}\sigma^{y}\Psi_{R_{i}+x})+h.c.]  \nonumber \\
&&-t\sum_{p_{n},R_{n}}(C_{p_{n}\sigma}^{\dag}C_{R_{n}\sigma}+h.c.)
\label{eq:hamiltonian}
\end{eqnarray}

Here $t=\frac{\hbar ^{2}}{2m^{\ast }a^{2}}$ is the hopping
parameter between two nearest-neighbor sites where $m^{\ast }$ is
effective mass of electrons and $a$ the lattice constant in the
2DEG bar. $\Psi _{R_{i}}^{\dag }=(C_{R_{i},\uparrow
},C_{R_{i},\downarrow })$ is the spinor vector for the site
$R_{i}$ ( $R_{i}=(x,y)$) in the Rashba SO system. $\sigma
^{x},\sigma ^{y}$ are the standard pauli matrices. $C_{p_{j}\sigma
}$ is the
annihilation operator for the site $p_{j}$ with spin $\sigma $ in lead p. $%
p_{n}$ and $R_{n}$ stand for nearest-neighbor pair on the two sides across
the interface of the SO system and the lead $p$. $w_{R_{i}}$ is the on-site
energy in the SO system, which can readily incorporate random impurity
potential. In a pure system we always set $w_{R_{i}}$ to zero. $t_{R}$ is
Rashba coupling coefficient.

Our calculations will follow the same spirit of the usual Landauer-B\"{u}%
ttiker's approach. But unlike the frequently used Green's function formalism%
\cite{LShengPRL2005,JLi2005}, in the formalism used in the present paper we
will solve the scattering wave functions in the whole system explicitly. To
this end, we firstly consider the scattering wave of an electron incident
from a lead. The real space wave function of an incident electron with spin $%
\sigma $ will be denoted as $e^{-ik_{m}^{p}x_{p}}\chi _{m\sigma }^{p}(y_{p})$%
, where $\chi _{m\sigma }^{p}(y_{p})$ denotes the $m$'th transverse mode
with spin index $\sigma $ in the lead $p$ and $k_{m}^{p}$ the longitudinal
wave vector. We adopt the local coordinate scheme for all leads. In the
local coordinate scheme, the longitudinal coordinate $x_{q}$ in lead $q$
will take the integer numbers $1,2,$...,$\infty $ away from the 2DEG
interface and the transverse coordinate $y_{q}$ take the value of $%
-N_{q}/2,...,N_{q}/2$. The longitudinal wave vector $k_{m}^{p}$ satisfies
the relation $-2t\cos (k_{m}^{p})+\varepsilon _{m}^{p}=E$, where $%
\varepsilon _{m}^{p}$ is the eigen-energy of the $m$'th transverse mode in
lead $p$ and $E$ the energy of the incident electron. Including both the
incident and reflected waves, the total wave function in the lead $q$ has
the the following general form:
\begin{eqnarray}
\psi _{\sigma \prime }^{pm\sigma }(x_{q},y_{q}) &=&\delta _{pq}\delta
_{\sigma \sigma \prime }e^{-ik_{m}^{p}x_{p}}\chi _{m\sigma }^{p}(y_{p})
\nonumber \\
&&+\sum_{n\in {q}}\phi _{qn\sigma ^{\prime }}^{pm\sigma
}e^{ik_{n}^{q}x_{q}}\chi _{n\sigma ^{\prime }}^{q}(y_{q})
\label{eq:swavefunction}
\end{eqnarray}%
where $\phi _{qn\sigma ^{\prime }}^{pm\sigma }$ stands for the scattering
amplitude from the ($m\sigma $) mode in lead $p$ to the ($n\sigma ^{\prime }$%
) mode in lead $q$. The scattering amplitudes $\phi _{qn\sigma ^{\prime
}}^{pm\sigma }$ will be determined by solving the Schr\"{o}dinger equation
for the entire system, which has now a lattice form and hence there is
\textit{a separate equation} for each lattice site and spin index. Since Eq.(%
\ref{eq:swavefunction}) is a linear combination of all out-going modes with
the same energy $E$, the Schr\"{o}dinger equation is satisfied automatically
in lead $q$, except for the lattice sites in the \emph{first row }( i.e., $%
x_{q}=1$ ) of the lead that are connected directly to the 2DEG bar. The wave
function in the first row of\emph{\ }a lead, which are determined by the
scattering amplitudes $\phi _{qn\sigma ^{\prime }}^{pm\sigma }$, must be
solved with the wave function in the 2DEG bar simultaneously due to the
coupling between the lead and the 2DEG bar. To simplify the notations, we
define the wave function in the 2DEG bar as a column vector $\psi $ whose
dimension is $2N$ ( $N$ is the total number of lattice sites in the 2DEG bar
). The scattering amplitudes $\phi _{qn\sigma ^{\prime }}^{pm\sigma }$ will
be arranged as a column vector $\phi $ whose dimension is $2M$ ( $M$ is the
total number of lattice sites in the first row of the leads ). From the
lattice form of the Schr\"{o}dinger equations for the 2DEG bar and the first
row of a lead, one can obtain the following equations reflecting the mutual
influence between the two parts:
\begin{eqnarray}
\mathbf{A}\psi &=&\mathbf{b}+\mathbf{B}\phi ,  \nonumber \\
\mathbf{C}\phi &=&\mathbf{d}+\mathbf{D}\psi ,  \label{eq:three}
\end{eqnarray}%
where $\mathbf{A}$ and $\mathbf{C}$ are two square matrices with a dimension
of $2N\times 2N$ and $2M\times 2M$, respectively; $\mathbf{B}$ and $\mathbf{D%
}$ are two rectangular matrices describing the coupling between the leads
and the 2DEG bar, whose matrix elements will depend on the actual form of
the geometry of the system. The vectors $\mathbf{b}$\textbf{\ }and\textbf{\ }%
$\mathbf{d}$ describe the contributions from the incident waves. Some
details of deduction as well as the elements of these matrices and vectors
have been given elsewhere\cite{Jiang2}.

After obtaining all scattering amplitudes $\phi _{qn\sigma ^{\prime
}}^{pm\sigma }$ , we can calculate the charge current in each lead through
the Landauer-Buttiker formula, $I_{p}=(e^{2}/h)\sum_{q}\sum_{\sigma
_{1},\sigma _{2}}(T_{p\sigma _{2}}^{q\sigma _{1}}V_{q}-T_{q\sigma
_{1}}^{p\sigma _{2}}V_{p})$, where $V_{q}=\mu _{q}/(-e)$ is the voltage
applied in the lead $q$ and $\mu _{q}$ is the chemical potential in the lead
$q$, $T_{q\sigma ^{\prime }}^{p\sigma } $ are the transmission probabilities
defined by $T_{q\sigma ^{\prime }}^{p\sigma }=\sum_{m,n}|\phi _{qn\sigma
^{\prime }}^{pm\sigma }|^{2}v_{qn}/v_{pm}$ and $v_{pm}=2t\sin (k_{m}^{p})$
is the velocity for the $m$'th mode in the lead $p$.

With the wave function $\psi _{\sigma ^{\prime }}^{pm\sigma }(R_{i})$ in the
2DEG strip at hand, the non-equilibrium spin density in the 2DEG strip can
also be calculated readily by taking proper ensemble average following
Landauer's spirit\cite{datta}. We assume that the reservoirs connecting the
leads at infinity will feed one-way moving particles to the leads according
to their own chemical potential. Following the Landauer's spirit\cite{datta}%
, let's normalize the scattering wave function $\psi ^{pm\sigma }$ so that
there is one particle for each incident wave, i.e., we normalize $%
e^{-ik_{m}^{p}x_{p}}\chi _{m\sigma }^{p}(y_{p})$ to $e^{-ik_{m}^{p}x_{p}}%
\chi _{m\sigma }^{p}(y_{p})/\sqrt{L}$ , where $L\rightarrow \infty $ is the
length of lead $p$. Meanwhile, the density of states in lead $p$ is $\frac{L%
}{2\pi }\frac{dk}{dE}=\frac{L}{2\pi \hbar v_{pm}}$, where $v_{pm}=2t\sin
(k_{m}^{p})$ is the velocity of the $m$'th transverse mode in lead $p$.
Adding the contributions of all incident channels of lead $p$ with
corresponding density of states ( DOS ), we can obtain the total
non-equilibrium spin density. In the linear transport regime, the spin
density can be calculated with the incident energy at the Fermi surface as:
\begin{equation}
\langle \vec{S}_{\alpha }(R_{i})\rangle =\frac{1}{2\pi }\sum_{pm\sigma }\mu
_{p}/\hbar v_{pm}\sum_{\sigma ^{\prime },\sigma ^{\prime \prime }}\psi
_{\sigma ^{\prime }}^{pm\sigma \ast }(R_{i})\vec{\sigma}_{\sigma ^{\prime
},\sigma ^{\prime \prime }}^{\alpha }\psi _{\sigma ^{\prime \prime
}}^{pm\sigma }(R_{i}),  \label{linear}
\end{equation}%
where $\langle \vec{S}_{\alpha }(R_{i})\rangle $ denotes the spin density at
a lattice site $R_{i}$ in the 2DEG strip and $\mu _{p}$ is the chemical
potential of lead $p$.

\section{Results and discussions}


When the system is in the equilibrium state, there will be no net spin
density since the Hamiltonian has time-reversal ( $T$ ) symmetry. However,
as a charge current flows from lead 1 to lead 2, non-equilibrium spin
density may emerge in the whole system, both inside the Rashba bar and in
the leads. In this section we will investigate the non-equilibrium spin
density under the condition of a fixed longitudinal charge current density.
In our calculations we will take the typical values of the electron
effective mass $m^{\ast }=0.04m_{e}$ and the lattice constant $a=3nm$\cite%
{NIttaPRL1997}. The chemical potential difference between the two leads will
be set by fixing the longitudinal charge current density to the experimental
value ( $\approx 100\mu A/1.5\mu m$ ) as reported in Ref\cite%
{WunderlichPRL2005} and the Fermi energy of the 2DEG bar will be set to $%
E_{f}=-3.8t$ throughout the calculations. We will limit our discussions to
the linear transport regime at zero temperature.

\begin{figure}[tbh]
\includegraphics[width=7cm]{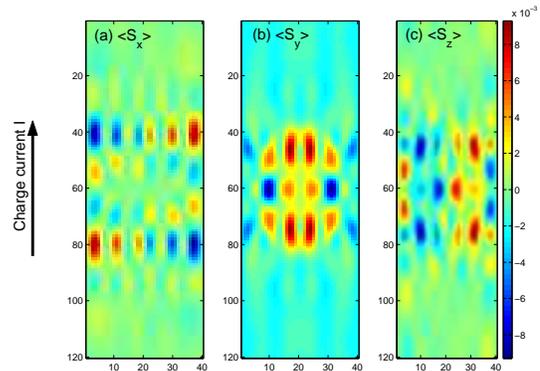}
\caption{(Color online)Typical pattern of current-induced non-equilibrium
spin density(in units of $\frac{h}{4\protect\pi}$) in a two terminal
structure. The lattice size for the central SO bar is $40\times40$. The
Rashba coupling strength is taken as $t_R$=0.1t. }
\label{pattern}
\end{figure}

To show clearly the characteristics of the current-induced spin polarization
in a two-terminal structure, below we will study the spin density in the
Rashba bar and in the leads separately. Firstly we study the spin density in
the Rashba bar. The typical spin density pattern obtained in a two terminal
structure is shown in Fig.\ref{pattern}. From the figures one can see that
the spatial distribution of the spin density inside the Rashba bar exhibits
some apparent symmetry properties, which are summarized in Eq.(\ref%
{eq:symmetryR}) given below. Theoretically speaking, these
symmetry properties are the results of some symmetry operations
implicit in the two-terminal problem. Let us explain this point in
some more detail. From the symmetry property of the Hamiltonian
(\ref{eq:hamiltonian}) and the rectangular geometry of the system
shown in Fig.\ref{fig:geometry}, one can see that in our problem
we have two symmetry operations $i\sigma _{y}P_{x}$ and $i\sigma
_{x}P_{y}$, where $P_{x}$ and $P_{y}$ denote the spacial
reflection manipulation $y\rightarrow -y$ and $x\rightarrow -x$
respectively and $i\sigma _{x}$ and $i\sigma _{y}$ the spin
rotation manipulation with the angle $\pi $ around the $S_{x}$ and
$S_{y}$ axes in spin space respectively. By considering these
symmetry operations, from Eq.(4) one can obtain immediately the
following symmetry relations\cite{Jiang2}:
\begin{eqnarray}
<S_{x,z}(x,y)>_{I} &=&-<S_{x,z}(x,-y)>_{I}  \nonumber \\
<S_{y}(x,y)>_{I} &=&<S_{y}(x,-y)>_{I}  \nonumber \\
<S_{x}(x,y)>_{I} &=&-<S_{x}(-x,y)>_{I}  \nonumber \\
<S_{y,z}(x,y)>_{I} &=&<S_{y,z}(-x,y)>_{I},  \label{eq:symmetryR}
\end{eqnarray}%
where $<\cdots >_{I}$ stands for the spin density induced by a longitudinal
charge current $I$ flowing from lead 1 to lead 2. The first two lines of Eq.(%
\ref{eq:symmetryR}) result from the symmetry operation $i\sigma _{y}P_{x}$,
which has been known widely before\cite{SQShen2004,Zyang2005}. The second
two, however, are results from the symmetry operation $i\sigma _{x}P_{y}$
and the $T$-reversal operation together\cite{Jiang2}. Firstly, due to the $T$%
-reversal invariance, the equilibrium spin density vanishes when
all chemical potentials are equal in Eq.(\ref{linear}), i.e.,
$<S_{\alpha
}(x,y)>_{eq}=<S_{\alpha }(x,y)>_{I}+<S_{\alpha }(x,y)>_{-I}=0$, where $%
<S_{\alpha }(x,y)>_{-I}$ denotes the spin density induced by a longitudinal
charge current $I$ from lead 2 to lead 1. Secondly, due to the geometry
symmetry of the system under the manipulation $i\sigma _{x}P_{y}$, we have $%
<S_{x}(x,y)>_{I}=<S_{x}(x,y)>_{-I}$ and $%
<S_{y,z}(x,y)>_{I}=-<S_{y,z}(x,y)>_{-I}$. Combining these two results, we
get the last two lines in Eq.(\ref{eq:symmetryR}).

\begin{figure}[tbh]
\includegraphics[width=7cm]{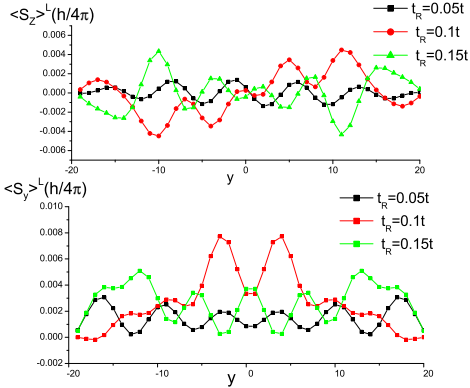}
\caption{(Color online)Non-equilibrium spin density averaged in longitudinal
direction in SO system , $<S_{y,z}>^{L}$, as a function of transverse
position, y. The y and z components has different reflection symmetry with
respect to x axis. The sign near the boundaries can be flipped by tuning
Rashba coupling strength $t_{R}$. }
\label{fig:accumulation}
\end{figure}

In Fig.\ref{fig:accumulation} we plot the longitudinally averaged spin
density inside the Rashba bar as a function of the transverse position $y$,
where $<\vec{S}(y)>^{L}$denotes the longitudinal averaged value of $<\vec{S}%
(x,y)>$ along the $x$ direction ( i.e., the direction of the charge current
flow ). Due to the symmetry relations shown in Eq.(\ref{eq:symmetryR}), the
non-vanishing components are $<S_{y,z}(y)>^{L}$. Moreover, under the spatial
reflection manipulation $y\rightarrow -y$, $<S_{y}(y)>^{L}$ is even and $%
<S_{z}(y)>^{L}$ is odd.
The fact that the out-of-plane component $<S_{z}(y)>^{L}$ has opposite signs
near the two lateral edges is consistent with the phenomenology of SHE%
\cite{Nikolic2005,WunderlichPRL2005}. But for a two-terminal lattice
structure with general lattice sizes, our results show that the spin density
does \emph{not} always develop peak structures near the two boundaries but
oscillate across the transverse direction. This is somehow different from
the naive picture of spin accumulation near the boundaries due to a spin
Hall current. The in-plane spin polarization is not related to the
phenomenology of spin Hall effect. It can be regarded as a general
magnetoelectric effect due to spin-orbit coupling\cite{HuangHu}. It should
be noted that, from the theoretical points of view, the relationship between
spin current the induced spin polarization is actually a much subtle issue
and is currently still under intensive debates. In this paper, however, we
will free our discussions from such controversial issues.

It is evident from Fig.~\ref{fig:accumulation} that not only the magnitude,
but also the sign of the spin accumulation near the two transverse
boundaries can be changed by tuning the Rashba coupling strength. This is
also reported in Ref\cite{Zyang2005} for a continuum two-terminal model.
Since the Rashba coupling strength can be tuned experimentally through gate
voltage, such a flipping behavior for spin accumulation might provide an
interesting technological possibility for the electric control of the spin
degree of freedom.


\begin{figure}[tbp]
\includegraphics[width=7cm]{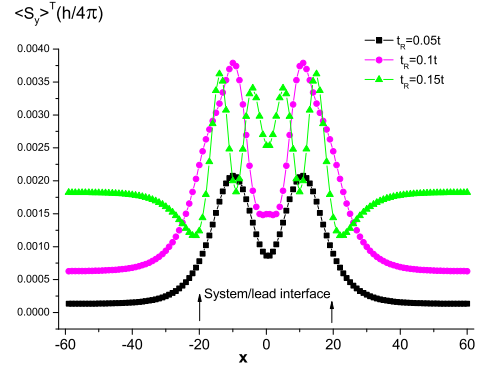}
\caption{(Color online)The variation of spin density averaged in transverse
direction of lead 2 versus Rashba coupling constant.}
\label{fig:polarization1}
\end{figure}

Now we present the most important theoretical prediction of this paper,
i.e., the charge-current induced spin polarization effect in the contacted
leads. Due to the first two lines in Eq.(\ref{eq:symmetryR}), after taking
average in the transversal direction ( which will be denoted by $<S_{\alpha
}>^{T}$ as a function of the longitudinal coordinate $x_{2}$ ), only $%
<S_{y}>^{T}$ will remain nonzero. In Fig.~\ref{fig:polarization1} $%
<S_{y}>^{T}$ is plotted versus the longitudinal coordinate $x$ of
the whole system. The lattice size of the whole system is set to
$120\times 40$, while the central $40\times 40$ lattice sites
represents the Rashba SO bar (see Fig.\ref{fig:geometry}). As
indicated clearly in the figure, the border between the Rashba bar
and the contacted leads are at $x=\pm 20$. The Rashba SO coupling
strength is chosen to be $t_{R}/t=0.05,0.1,0.15$. As can be seen
from Fig.~\ref{fig:polarization1}, $<S_{y}>^{T}$ changes with the
coordinate $x$ both inside the Rashba bar and in the leads, but in
the leads it changes with the coordinate $x$ only in a much narrow
region close to the border between the leads and the Rashba bar
and will reach to a fixed value further away from the border. This
phenomenon is the \emph{lead spin polarization effect} in this
paper. Theoretically, this phenomenon can be understood as
following. For an incident wave $(m,\sigma )$ with a Fermi energy
$E_{f}$ in lead 1, the scattering wave function in lead 2 is:
$\psi _{\sigma ^{\prime }}^{1m\sigma }(x_{2},y_{2})=\sum_{n}\phi
_{2n\sigma ^{\prime }}^{1m\sigma }e^{ik_{n}x_{2}}\chi
_{n}(y_{2})$, where $-2t\cos (k_{n})+\varepsilon _{n}=E_{f}$. Then
the spin density $<\vec{S}(x_{2},y_{2})>$ can be calculated as (
dropping off a normalization factor ),
\[
\sum_{m,\sigma }\frac{1}{v_{1m}}\sum_{\alpha ,\beta }\psi _{\alpha
}^{1m\sigma \ast }(x_{2},y_{2})\vec{\sigma}_{\alpha ,\beta }\psi _{\beta
}^{1m\sigma }(x_{2},y_{2}),
\]%
Thus, after taking average along the transversal direction, we have
\begin{eqnarray}
\lefteqn{<\vec{S}(x_{2})>^{T}=(1/N_{2})\sum_{y_{2}}<\vec{S}(x_{2},y_{2})>}
\nonumber \\
&\propto &\sum_{m,\sigma }\frac{1}{v_{1m}}\sum_{n,\alpha ,\beta }[\phi
_{2n\alpha }^{1m\sigma }e^{-ik_{n}x_{2}})^{\ast }\vec{\sigma}_{\alpha ,\beta
}(\phi _{2n\beta }^{1m\sigma }e^{-ik_{n}x_{2}})],  \label{eq:ferro}
\end{eqnarray}%
where we have used the orthogonality relation for the transverse modes: $%
\sum_{y_{2}}\chi _{n}(y_{2})\chi _{n^{\prime }}(y_{2})=\delta
_{nn^{\prime }} $. Evidently, the summand $[\cdots ]$ in
Eq.~(\ref{eq:ferro}) is independent of $x_{2}$ as long as $k_{n}$
is real ( i.e., for those longitudinal propagating modes ). If
$k_{n}$ is imaginary, the corresponding mode will describe an
exponentially localized state in lead 2 in the vicinity of the
border between the lead\emph{\ }and the Rashba bar. Such
evanescent components\cite{Usaj2005} can contribute to the local
charge density and spin density only in the vicinity of the
border. At some distance far away
from the interface, the contribution of the evanescent modes to $<\vec{S}%
(x_{2})>^{T}$will decay exponentially as $x_{2}$ increase and hence $<\vec{S}%
(x_{2})>^{T}$ will be independent of $x_{2}$, as illustrated in Fig.~\ref%
{fig:polarization1}. \ This result implies that the spin
polarization inside the Rashba bar can be inducted out by the
driving charge current to the leads and manifest itself in an
amplified way, i.e., the leads will become spin polarized and thus
an electric-controllable spin state in the leads can be realized.
Such a lead spin polarization effect might be more easily
detectable by magnetic or optical methods than the spin
polarization inside the mesoscopic Rashba bar.

\begin{figure}[tbp]
\includegraphics[width=7cm]{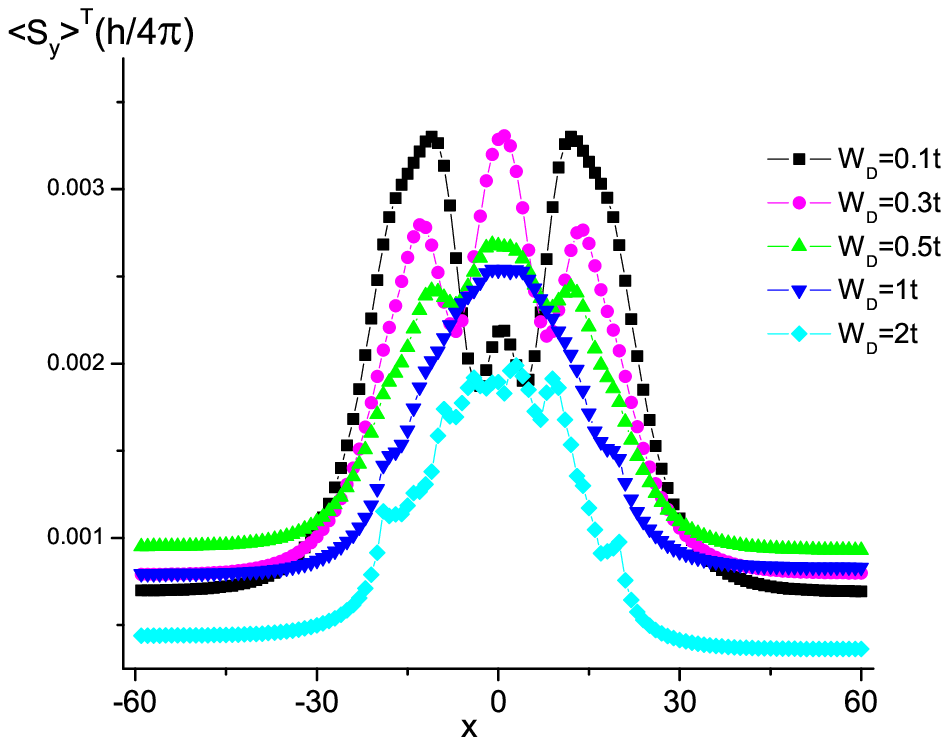
}
\caption{(Color online)The variation of spin density averaged in
transverse direction \emph{versus} longitudinal coordinate at the
presence of disorder scattering. } \label{fig:polarization2}
\end{figure}

The results presented above is obtained in the absence of impurity
scattering. To simulate spin-independent disorder scattering, we
assume a uniformly distributed random potential $w\in \lbrack
-W_{D},W_{D}]$ in the Rashba bar with a disorder strength $W_{D}$.
The non-equilibrium spin density can be obtained by taking average
for a number of random realizations of local potentials. We
averaged 1000 random realizations in
all calculations. In Fig.~\ref{fig:polarization2} we show how the curves of $%
<S_{y}>^{T}$ versus $\ x$ change with $W_{D}$. All curves are calculated
with $t_{R}=0.1t$. Noticeably, inside the Rashba bar the height of $%
<S_{y}>^{T}$decreases with increasing $W_{D}$ but the lead spin
polarization will firstly \emph{increase} in the weak disorder
regime ( $W_{D}<0.5t$ ) and then decrease in the stronger disorder
regime ( note that our calculations are carried out under the
condition of a fixed longitudinal charge current density). Thus,
there exists a certain regime in which disorder  can
\emph{enhance} the saturation value of the spin polarization in
the  leads. The slight asymmetric form of $W_{D}=2t$ line means
that we need to take more random configurations of local potential
for large $W_{D}$ value, which is reasonable. Of course, there are
some other factors such as spin-orbit coupling or impurity
scattering in the leads that may reduce or suppress the lead spin
polarization effect. Further quantitative study is needed in order
to clarify the role of these factors.

\section{Conclusion}

To conclude, in this paper we have presented a theoretical study on the
non-equilibrium lead spin polarization effect in a two-terminal mesoscopic
Rashba bar under the condition of a fixed longitudinal charge current
density. We have predicted that a finite amount of non-equilibrium spin
polarizations can be induced in the leads by the spin-orbit coupling inside
the mesoscopic Rashba bar when the longitudinal charge current circulate
through it. Such a lead spin polarization effect can survive in the presence
of weak disorder inside the Rashba bar and thus might be observable in real
experiment. 
Such an effect might provide a new kind of electric-controllable
spin state which is technically attractive. But it should be
stressed that, for real systems, due to the existence of disorder
scattering or spin-orbit interactions or other spin decoherence
effects in the leads, the lead spin polarization effect predicted
in the present paper may be weakened. These factors need to be
clarified by more detailed theoretical investigations in the
future.

\begin{acknowledgments}
The author is grateful to Research Center for Quantum Manipulation
in Fudan University for hospitality during his visit in summer of
2005. He would like to thank T.Li, R.B.Tao, S.Q.Shen, Z.Q.Yang,
L.B.Hu for various helpful discussions and L.B.Hu for help on
polishing the manuscript. This work was supported by Natural
Science Fundation of Zhejiang province ( Grant No.Y605167 ) and
Research fund of ZheJiang normal university.
\end{acknowledgments}

\bibliography{leadpolarization}

\end{document}